\documentclass[prb,twocolumn,showpacs,preprintnumbers,amsmath,amssymb]{revtex4}
\usepackage{graphicx}
\usepackage{dcolumn}
\usepackage{bm}

\usepackage{amssymb}
\usepackage{epstopdf}
\newcommand{\be}{\begin{equation}}
\newcommand{\ee}{\end{equation}}
\newcommand{\bea}{\begin{eqnarray}}
\newcommand{\eea}{\end{eqnarray}}

\newcommand{\tr}{{\rm tr\/}\,}

\newcommand{\lp}{\left(}
\newcommand{\rp}{\right)}

\renewcommand{\phi}{\varphi}
\renewcommand{\epsilon}{\varepsilon}
\renewcommand{\vec}[1]{{\bf #1}}

\newcommand{\Tr}{\mathrm{Tr}} 
\newcommand{\sign}{\mathrm{sign}}
\begin{document}

\title{Quantum Anomalous Hall State in Bilayer Graphene}
\author{Rahul Nandkishore and Leonid Levitov}
\affiliation{Department of Physics, Massachusetts Institute of Technology, Cambridge, MA02139}

\begin{abstract}
We present a symmetry-based analysis of competition between different gapped
states that have been proposed in bilayer graphene (BLG), which are all
degenerate on a mean field level.
We classify
the states in terms of a hidden SU(4) symmetry, and distinguish symmetry
protected degeneracies from accidental degeneracies. One of the states, which
spontaneously breaks discrete time reversal symmetry but no continuous
symmetry, is identified as a Quantum Anomalous Hall (QAH) state, which exhibits
quantum Hall effect at zero magnetic field. We investigate the lifting of the
accidental degeneracies by thermal and zero point
fluctuations, taking account of the modes softened under a renormalisation group procedure (RG). Working in a
`saddle point plus quadratic
fluctuations' approximation, we identify two types of RG-soft modes which have
competing effects. Zero point fluctuations, dominated by `transverse'
modes which are unique to BLG, favor the QAH state. Thermal
fluctuations, dominated by `longitudinal'
modes, favor a SU(4) symmetry breaking multiplet of states. We
discuss the phenomenology and experimental signatures of the QAH state in BLG,
and also propose a way to induce the QAH state using weak external magnetic
fields.
\end{abstract}

\maketitle

\section{Introduction}

The Quantum Anomalous Hall (QAH) insulator is a state of matter where spontaneous breaking of time reversal symmetry produces (integer) quantum Hall effect in the absence of any external magnetic field. First predicted in 1988 \cite{Haldane}, the QAH state has never yet been observed. 
In the recent literature on interaction driven topological insulators \cite{Shou-Cheng, Kivelson}, 
the elusiveness of the QAH state has been ascribed to fluctuations, which 
typically disfavor the QAH state with respect to a Quantum Spin Hall (QSH) state, which is degenerate with the QAH state on a mean field level. 
Here we point out that the fluctuations which govern the competition of different gapped phases proposed in bilayer graphene (BLG) \cite{Min, Nandkishore, Zhang10} are dominated by the modes not present in the models \cite{Shou-Cheng, Kivelson}, leaving 
open the door to formation of a QAH state at zero field in BLG. Also, we will propose a mechanism for 
inducing the QAH state using external fields.



The theoretical literature on BLG predicts instabilities to numerous strongly correlated states, which are gapped \cite{Min, Nandkishore, Zhang10} or gapless \cite{Yang} depending on the way the electron-electron interaction is modeled. The numerous gapped states predicted in the literature are all degenerate at the level of mean field theory \cite{Min, Nandkishore}, and have the same instability threshold under one loop RG \cite{Zhang10}. The relation between these different states, and their experimental signatures, have not yet been understood. 

Meanwhile, recent experiments indicate that the gapped state observed in charge neutral BLG in quantizing magnetic fields \cite{Feldman} persists down to low fields, crossing over to another gapped state at zero field \cite{Feldman2}. However, the nature of the gapped state at zero field is unknown. 
Hence, clarifying the relation between different gapped states and understanding their physical properties is an interesting and timely task.

Here we present a unifying symmetry based analysis of strongly correlated states in BLG. The states predicted in Refs.[\onlinecite{Min, Nandkishore, Zhang10}] are classified according to a hidden SU(4) flavor symmetry into symmetry breaking multiplets and an SU(4) invariant singlet. The SU(4) singlet is a QAH state. The degeneracy of the multiplets and the singlet is an artefact of the approximations made in the analysis,
and will be lifted upon taking fluctuation effects into account. 

Our analysis of fluctuations in BLG focuses on the effect of the modes softened under RG. Those include the `longitudinal' fluctuation modes (L-modes) analogous to those discussed in Ref.[\onlinecite{Shou-Cheng, Kivelson}], and also `transverse' fluctuation modes (T-modes) which are unique to BLG. We find that these two types of modes have competing effects: while the L-modes favor the symmetry breaking multiplets, the T-modes favor the SU(4) invariant QAH state. 
The zero-point fluctuations are dominated by the 
T-modes, and hence appear to favor 
a QAH state at zero temperature. Meanwhile, thermal fluctuations are dominated by the L-modes,
and favor the symmetry breaking multiplets. We speculate that thermal fluctuations may drive a
phase transition from the QAH state at low temperatures to a SU(4) symmetry breaking state at higher temperatures, and estimate the transition temperature. We also discuss the phenomenology of the QAH state, its possible experimental signatures, and 
propose a way to further stabilize it using external magnetic fields. 


\section{SU(4) symmetry}

In this section we show that within the often-used approximation where the difference between interlayer and intra-layer interactions is neglected \cite{Min, Nandkishore, Zhang10, Barlas, Nilsson}, the interacting Hamiltonian is invariant under rotations in a suitably defined four-dimensional flavor subspace.
Specifically, we perform a unitary transformation by exchanging the sublattices $A$ and $B$ in one of the valleys, upon which the single particle Hamiltonian becomes identical for all spin and valley species, while the layer and sublattice blind interactions are left unchanged.

Before entering the discussion of the SU(4) invariance in BLG, we recall that electronic states in BLG at low energy are described by wave-functions 
on the $A$ and $B$ sublattices of the upper and lower layers
\cite{Novoselov}, and are fourfold degenerate in spin and valley. 
To analyze the structure of the Hamiltonian,
it will be convenient to combine the spin and valley components in a single eight-component wavefunction  $\psi_{\alpha, s, v}(\vec{x})$, where $\alpha$ is the sublattice (layer) index. We shall use the Pauli matrices in sublattice, spin and valley space, denoted below by  $\tau_i$, $\sigma_i$ and $ \eta_i$, respectively. The low energy non-interacting Hamiltonian may then be written as 
\be
\label{eq: Hamiltonian}
H_0 = \frac{(p_x + i p_y \eta_3)^2}{2m} \tau_- + \frac{(p_x - ip_y \eta_3)^2}{2m} \tau_+  
,
\ee
where $ \tau_{\pm} = \tau_1 \pm i \tau_2$. Here $m = 0.05 m_e$ is the effective mass. Because of the presence of $ \eta_3$ in Eq.(\ref{eq: Hamiltonian}), the single particle Hamiltonian is not invariant under rotations of valley components. To bring it to an SU(4) invariant form,  
we perform
a unitary transformation on all operators 
\be\label{eq: transformation}
\tilde O = U O U^{\dag}
,\quad 
U = \frac{1+ \eta_3}{2} + \frac{1 - \eta_3}{2} \tau_1 
.
\ee
This transformation does not act on the spin space, however it mixes the layer and valley indices of the wavefunction $\psi_{\alpha, s, v}(\vec{x})$ by interchanging the $\tau$ pseudospin component (layer) in one of the valleys. As a result, $\tau_+$ and $\tau_-$ are interchanged and $\tau_3$ changes sign in the $\eta_3 = -1$ valley, after which the free-particle Hamiltonian, Eq.(\ref{eq: Hamiltonian}), becomes identical in both valleys.

Defining $p_{\pm} = p_x \pm i p_y$, the transformed non-interacting Hamiltonian takes the compact form 
%
\be
H_0 = \frac{p_+^2}{2m} \tilde \tau_- + \frac{p_-^2}{2m} \tilde \tau_+ 
,
\label{eq: transformed Hamiltonian}
\ee
where $\tilde \tau_+$ and $\tilde \tau_-$ are obtained by transforming $\tau_+$ and $\tau_-$ according to Eq.(\ref{eq: transformation}). This single particle Hamiltonian is manifestly invariant under $SU(4)$ rotations in the transformed spin/valley flavor space.

Meanwhile, electron interactions can be described by a many-body Hamiltonian written in terms of $\rho_{\vec{q}} = \sum_{\vec{p}} \psi^{\dag}_{\vec{p}} \psi_{\vec{p} + \vec{q}}$ (the density summed over layers) and $\lambda_{\vec q} =  \sum_{\vec{p}} \psi^{\dag}_{\vec{p}} \tilde \tau_3 \tilde \eta_3 \psi_{\vec{p} + \vec{q}}$ (the density difference between layers). The interacting Hamiltonian, which incorporates a difference between interlayer and intralayer interaction \cite{Nilsson}, can be written as
\be\label{eq:Hint}
H =\sum_{\vec{p}} \psi^{\dag}_{\vec{p}} H_0 
 \psi_{\vec{p}} 
+\frac12 \sum_{\vec{q}} V_+ (q) \rho_{\vec{q}} \rho_{-\vec{q}} + V_- \lambda_{\vec{q}}\lambda_{-\vec{q}},
\ee
where $V_+(q) = 2\pi e^2 /\kappa q$ is the Coulomb interaction, and $V_- = \pi e^2 d/\kappa$ accounts for the layer polarization energy (here $d=3.5 \AA$ is the BLG layer separation). The $\rho\rho$ term,
which is isotropic in flavor space and thus is SU(4) invariant, dominates because $d$ is small compared to
\be\label{Bohrs_radius}
a_0 = \hbar^2\kappa/me^2 = 10 \kappa\, {\rm \AA}
,
\ee
the characteristic lengthscale set by interactions \cite{Nandkishore}. We therefore approximate by neglecting $V_-$, an approximation that becomes exact in the weak coupling limit, where $d/a_0 \rightarrow 0$. Under this approximation, the Hamiltonian is invariant under SU(4) flavor rotations, generated by the operators $\tilde \eta_i$ and $\tilde \sigma_i$. We will henceforth drop the $\sim$  symbols for notational convenience, and will refer to the operators $\tilde \tau$, $\tilde \eta$ and $\tilde \sigma$ as $\tau$, $\eta$ and $\sigma$ respectively. All operators are assumed to be transformed operators unless specified otherwise.


\section{Classification of states and topological properties}

In the transformed basis, the mean field Hamiltonian for the gapped states described in \cite{Min, Nandkishore, Zhang10} may be written as
\be \label{eq: full Hamiltonian}
H = 
\frac{p_+^2 \tau_- +p_-^2  \tau_+}{2m}
+ \Delta  \tau_3 Q 
,
\ee
where $m = 0.05 m_e$ is the effective mass. Here the Pauli matrices $ \tau_i$ act on the transformed sublattice space, and $Q$ is a $4\times4$ hermitian matrix in the transformed spin-valley space (flavor space), satisfying $Q^2 = 1$.

Since unitary hermitian matrices have eigenvalues $\pm1$, all gapped states can be classified as ($M_>$, $M_<$), where $M_>$ and $M_<$ are the numbers of $+1$ and $-1$ eigenvalues of $Q$ respectively. There are three general types of states:  $(2,2)$, $(3,1)$, and $(4,0)$. There is an additional $Z_2$ symmetry associated with the overall sign of $Q$ which is absorbed into the sign of $\Delta$. Following Refs.[\onlinecite{Haldane,Jackiw}], the Hall conductance of a state ($M_>$, $M_<$) can be written as 
%
\begin{equation}\label{eq: hall conductance}
\sigma_{xy} =  (M_> - M_<)\, 
\frac{e^2}{h},
\end{equation}
where we took into account an additional factor of $2$ due to the Berry phase $2\pi$ in BLG \cite{Novoselov}.
The (4,0) and (3,1) states, which have $M_> \neq M_<$, thus exhibit a quantized Hall conductance at zero magnetic field--the hallmark of a QAH state. 
Because these states have $\sigma_{xy} \neq 0$, they must spontaneously break time reversal symmetry. We will henceforth focus on comparing the (4,0) and (2,2) states, since the (3,1) states are intermediate between the two. We will refer to the (4,0) state as the QAH state, but it should be remembered that the (3,1) states are also QAH states. In contrast, the (2,2) states have $\sigma_{xy}=0$, and preserve time reversal symmetry, but instead exhibit quantum flavor Hall effect. If we parameterize the flavor space by Pauli matrices $\eta_i$ and $\sigma_i$ in transformed valley and spin space respectively, then the $Q =  \sigma_3$ state  is a QSH state (Ref.[\onlinecite{KaneMele}]),
while the $Q =  \eta_3$ state is a Quantum Valley Hall (QVH) state (Ref.[\onlinecite{Morpurgo}]). 

These states are analogs of the `topological Mott insulators' discussed in Refs.[\onlinecite{Shou-Cheng, Kivelson}], and as such host topologically protected edge states. The counter-propagating valley modes for the QVH state were worked out in Ref.[\onlinecite{Morpurgo}], the co-propagating charge modes and the counter-propagating spin modes for the QAH and QSH states follow similarly.
The protection of edge modes is strongest for the (4,0) state due to the  unidirectional, chiral character of these modes.
The counter-propagating spin currents in the QSH state are protected in the absence of spin-flip scattering, while the counter-propagating valley currents in the QVH state are protected in the absence of intervalley scattering (e.g. by short range disorder).

\begin{figure}
\includegraphics[width = 3.5in]{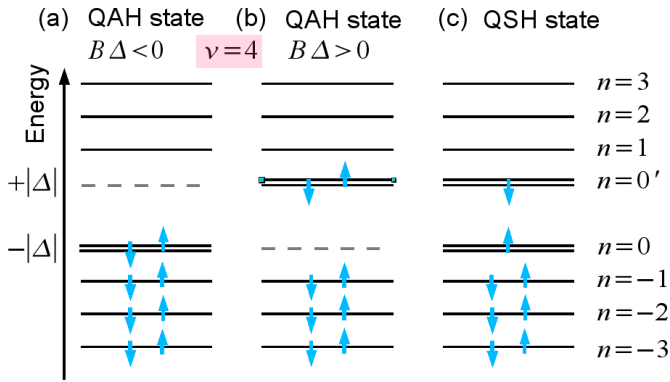}
\caption{Landau level spectrum of the QAH and QSH states. Note an anomalous Landau level in the QAH state that has no particle-hole-symmetric counterpart. Occupation of this anomalous Landau level allows the QAH state (a) to lower its energy relative to the states (b,c) at filling factor $\nu = 4$.}
\label{fig: phase diagram}
\vspace{-5mm}
\end{figure}

We note that the above classification of states superficially resembles that arising 
in an entirely different problem, namely the Quantum Hall Ferromagnet (QHF) in graphene in quantizing magnetic field\cite{Yang2}. In the latter case, however, the integers $M_>$ and $M_<$ are fixed by the electron density, i.e. by filling of the four-fold degenerate zeroth Landau level. In the QHF problem, spontaneous time-reversal symmetry breaking cannot occur: the analog of the (4,0) QAH state is a fully filled zeroth Landau level exhibiting quantized Hall conductance $2e^2/h$. Furthermore, in the QHF problem there is no competition between states with different $M_>$ and $M_<$ values, which is the main question of interest for us here. 

\section{Lifting accidental degeneracies using external fields}

The SU(4) symmetry of the Hamiltonian guarantees the degeneracy of all states within a given manifold ($M_>$, $M_<$), even when the states involved have very different physical properties. For example, the QVH state is a ferroelectric state which polarizes the layers by charge, while the QSH state polarizes the layers by spin and valley. Nonetheless, the two states  are related by SU(4) transformations, and are hence degenerate within the approximations leading to SU(4) symmetry.

In contrast, the degeneracy of the different manifolds $(M_>, M_<)$ is purely accidental, and may be lifted in the presence of a weak SU(4) invariant perturbation. As an example, we consider application of a weak transverse magnetic field $B$. Incorporated in the Hamiltonian (\ref{eq: full Hamiltonian}) through the replacement $\vec{p} \rightarrow \vec{p} - e\vec{A}$, it preserves the SU(4) symmetry, and causes the spectrum to split into Landau levels \cite{McCann} with an energy spacing of order $\hbar \omega_c$, where $\omega_c = eB/mc$.
The Zeeman energy $2 \mu_B B  \sigma_3$ is not $SU(4)$ invariant, but may be neglected since $\hbar \omega_c \gg 2\mu_B B$. When $\Delta = 0$ and Zeeman terms are neglected, the Landau level spectrum is particle-hole symmetric and is fourfold degenerate in flavors. 


Crucially, the $T$ non-invariance of the mass term $\Delta  \tau_3$ means that the Landau level spectrum for the (4,0) state is not invariant under $B \rightarrow -B$  and is not particle-hole symmetric \cite{Jackiw}. In particular, the zeroth Landau level, which has an additional two-fold orbital degeneracy \cite{McCann}, forms at energy $\Delta \sign B$ only, and has no counterpart at $-\Delta \sign B$ 
(see Fig.\ref{fig: phase diagram}a,b). This breaking of particle hole symmetry can be exploited to 
induce the (4,0) state using magnetic fields.

We illsutrate this by comparing the energies
of the $(4,0)$ QAH state and the $(2,2)$ QSH state in external magnetic field at filling factor $\nu = 4$.
In a magnetic field, 
these states are no longer degenerate, because of the anomalous Landau level. 
It is clear from Fig.\ref{fig: phase diagram} that the QAH state with the appropriate sign of $\Delta$ (such that $B\Delta <0$) is favored over the QSH state:
\begin{equation}\label{eq: mechanism}
F_{4,0} - F_{2,2} = \frac{4 \Delta B}{\Phi_0} < 0, 
\end{equation}
where $\Phi_0$ is the flux quantum and $F_{M_>,M_<}$ is the free energy per unit area for a state $(M_>,M_<)$. 
This mechanism for lifting the degeneracy between QAH and QSH states in favor of the QAH state applies to all systems where there is such a degeneracy, including the models studied in Refs.[\onlinecite{Shou-Cheng, Kivelson}]. Of course, at finite $B$, there is no time reversal symmetry, so the state realized is not a true QAH state, but rather is a state showing quantum Hall effect at anomalously low magnetic fields, which is smoothly connected to a QAH state at $B=0$.

The analysis above is valid only for sufficiently small $B$, when BLG at $\nu = 4$ is not far from charge neutrality. This is because the excitonic instability that generates the gap $\Delta$ (Refs.[\onlinecite{Min, Nandkishore, Zhang10}]) is suppressed by detuning away from charge neutrality. 

%
%
%

\section{Saddle point analysis}

We now investigate the energy splitting between the different manifolds at $B=0$ by going beyond a mean field approximation, and including the effect of fluctuations. We consider BLG in the presence of screened Coulomb interactions between electrons. A  static screening approximation, ignoring the effects of dynamical screening\cite{Nandkishore}, is sufficient to understand the main features.
In this approximation, the interaction is short range, and we can write the partition function as a functional field integral in Euclidean time, 
\be
Z = \int D\psi^{\dag} D\psi \exp\lp-\int dx \mathcal{L}[\psi^{\dag}(x), \psi(x)]
\rp ,
\ee
where $x = (t, \vec{r})$, $dx = dt d^2r$, the $\psi$ fields are fermionic fields, with the Lagrangian
%
\begin{equation}
\mathcal{L} = \psi^{\dag} (\partial_t +  H_0) \psi + \sum_{j,k=1...8}\frac{\lambda}{2} \psi^{\dag}_{j} \psi^{\dag}_{k}\psi_{k} \psi_{j}\label{eq: coupling}.
\end{equation}
Here $j$, $k$ are combined sublattice and flavor indices, 
and $H_0$ is the non-interacting Hamiltonian (given by Eq.(\ref{eq: full Hamiltonian}) at $\Delta = 0$). The coupling constant $\lambda$ represents the statically screened Coulomb interaction, which in the RPA model takes value \cite{DasSarma} $\lambda = 1/(4 \nu_0 \ln 4)$, where $\nu_0 = m/2\pi$ is the non-interacting single species density of states. 

We now decouple the four fermion interaction term via a Hubbard-Stratonovich transformation in the exchange channel, to obtain 
$Z = \int D\psi^{\dag} D\psi D h \exp[- \int dx \mathcal{L}(\psi^{\dag}, \psi, h)]$, where 
\begin{equation}
\mathcal{L} =\psi^{\dag} \big[\partial_t + H_0 + h \big] \psi + \frac{1}{2\lambda} \Tr [h h^{\dag}] \label{eq: lagrangian}.
\end{equation}
Here, $h$ is an $8\times8$ hermitian matrix, which we write as  $h = M \otimes Q$, where $M$ is a $2\times2$ hermitian matrix in sublattice space and $Q$ is a $4\times 4$ hermitian matrix in flavor space. The gapped states (Ref.[\onlinecite{Min, Nandkishore, Zhang10}]) correspond to taking $M = \Delta  \tau_3$. 
Integrating out the fermions yields $Z = \int D(Q) D(\Delta) \exp( - \int dx\mathcal{L}[\Delta(x), Q(x)]$, where 
\begin{equation}
\mathcal{L}(\Delta, Q) = - \Tr \ln \big[\partial_t + H_0 + \Delta  \tau_3 Q\big] + \frac{\Delta^2}{\lambda} \Tr [Q^2] \label{eq: effective action}.
\end{equation}
The SU(4) flavor invariance manifests itself in an exact SU(4) flavor degeneracy of the many body states. Upon minimizing the action (\ref{eq: effective action}) in a saddle point approximation, we find $Q^2 = 1$, and 
%
%
$\Delta= \Lambda \exp(- 2/\lambda \nu_0)$, where $\Lambda \approx 0.4 \,{\rm eV}$ is the bandwidth for the two band Hamiltonian. This gives
the mean field Hamiltonian, Eq.(\ref{eq: full Hamiltonian}). 

We note that instead of decoupling the interaction in the excitonic channel $h =\tau_3 \otimes Q$, we could have chosen the channel $h =\tau_{1,2} \otimes Q$. This choice would lead us to the nematic state of Ref.\onlinecite{Yang}, which is gapless, but breaks lattice rotation symmetry. However, the nematic state is higher in energy than the gapped states at the saddle point level, so we will concentrate on the gapped states, and specifically on the lifting of the accidental degeneracies by thermal and zero-point fluctuations. 

Our symmetry analysis, involving multiplets $(M_>,M_<)$ for different matrices $Q$, could also be applied to the nematic state \cite{Yang}. However, the fluctuation analysis 
cannot be perfomed because the $\tau_3\delta Q$ mode 
has negative rigidity, i.e. the nematic mean field is unstable.
 


%

\section{Lifting the degeneracy: zero-point fluctuations}
\label{sec: zero-point}

We first analyze the case of zero temperature, when the degeneracy is lifted by zero point fluctuations.
The most important fluctuation modes are those that are softened under RG. In BLG, this means the `L' modes $ \tau_3 \delta Q$, which describe fluctuations longitudinal with respect to the order parameter in sublattice space, and also the `T' modes $ \tau_{1,2} \delta Q$, which describe fluctuations transverse to the order parameter in sublattice space. In that, $\delta Q$ is an arbitrary $4\times 4$ hermitian matrix. 

We therefore expand the action in Eq.(\ref{eq: effective action}) to quadratic order in the fluctuation modes $\tau_{\alpha} \otimes \delta Q^{\alpha}$, $\alpha = 1,2,3$, to obtain 
\begin{equation}\label{eq: fluctuation expansion}
\delta^2 S = \sum_{ijkl\alpha\beta} \sum_{\omega, \vec{q}} \delta Q^{\alpha}_{ij, \omega \vec{q}} K^{\alpha \beta}_{ijkl}(\omega, \vec{q}) \delta Q^{\beta}_{k l, -\omega, -\vec{q}}. 
\end{equation}
Here, Latin indices $i,j,k=1...N$ refer to fermion flavor, whereas Greek indices $\alpha, \beta = 1,2,3$ refer to the Pauli matrices $\tau_{\alpha}$ that parameterize the fluctuations in sublattice space. 
The matrix $K$ is defined by 
\begin{equation}\label{eq: K}
K^{\alpha \beta}_{ijkl}(\omega, \vec{k}) = \delta_{il} \delta_{jk} \left(\frac{\delta_{\alpha \beta}}{\lambda} + \Pi^{\alpha \beta}_{ij} (\omega, \vec{k})\right)
,
\end{equation}
where we have introduced the polarization operator 
\be\label{eq: pi}
\Pi^{\alpha \beta}_{ij}(\omega, \vec{q})  = \int \frac{d^2 pd\epsilon}{2(2\pi)^3}
\Tr \lp \tau_{\alpha} G_{i}(p_+)\tau_{\beta} G_{j}(p_-) \rp 
.
\ee
It is convenient to choose a diagonal background state $Q = \zeta_i \delta_{ij}$, where $\zeta_i = \pm 1$, so that the Greens function takes a form diagonal in the flavor space,
\be
G_{i}(p_\pm)=\frac{1}{i(\epsilon \pm \frac12 \omega) - H_0(\vec{p} \pm \frac12 \vec{q}) - \zeta_i \Delta  \tau_3}
.
\ee
%
%
The trace in Eq.(\ref{eq: pi}) goes over sublattice indices, but not over flavors. 

The matrix $K$ is positive definite,
so we may integrate out fluctuations to obtain an expression for the fluctuation contribution to the free energy, 
%
\be
F_{\rm fluct} = \frac{1}{2} \sum_{\alpha  i j} \sum_{\omega \vec{k}} \ln K^{\alpha \alpha}_{i j ji}(\omega, \vec{k})
,
\ee
where we took into account that the only contribution comes from the diagonal terms,
$\alpha=\beta$, $i=l$, $j=k$. 
We now subtract the fluctuation energy of the $(4,0)$ QAH state from that of the $(2,2)$ state, to obtain
%
\be
\label{eq: deltaf}
\delta F = F_{\rm fluct,(4,0)} - F_{\rm fluct,(2, 2)}
= 4 \sum_{\alpha =1}^3 \ln \frac{\frac{1}{\lambda}+\Pi^{\alpha \alpha}_{>>}}{\frac{1}{\lambda} + \Pi^{\alpha \alpha}_{><}}
,
\ee
where $\Pi^{\alpha \alpha}_{>>}$ and $\Pi^{\alpha \alpha}_{><}$ are defined by Eq.(\ref{eq: pi}), with $(\zeta_i, \zeta_j) = (1,1)$ and $(1,-1)$ respectively:
%
\begin{eqnarray}\label{eq: pi>>}
&&\Pi^{\alpha \beta}_{>>}\!(\omega, \vec{q}) \!=  \!\!\! \int \!\!
\frac{d^2 pd\epsilon}{(2\pi)^3} 
 \frac{1}{2}\Tr \left( \tau_{\alpha} G_>(p_+)  \tau_{\alpha} G_>(p_-)\right) ,
\\\label{eq: pi><}
&& \Pi^{\alpha \beta}_{><}\!(\omega, \vec{q})  \!= \!\!\!  \int  \!\!
\frac{d^2 pd\epsilon}{(2\pi)^3} 
 \frac{1}{2} \Tr\left( \tau_{\alpha} G_>(p_+)  \tau_{\alpha} G_<(p_-)\right)
,
\\\nonumber
&& G_{>(<)}(\epsilon,\vec p) = \frac{1}{i\epsilon - H_0(\vec{p}) \mp \Delta  \tau_3}
,
\end{eqnarray}
where we used a shorthand notation $p_\pm=(\epsilon \pm \frac12 \omega, \vec{p} \pm \frac12 \vec{q})$.
%
To analyze the effect of competition of different modes in full detail, below we compare the fluctuation energy for the states of different type $(M_>,M_<)$.

To evaluate the difference of fluctuation energies, given by  Eq.(\ref{eq: deltaf}), it is convenient to rewrite it as
\be
\delta F = 4 \sum_{\alpha =1}^3 \ln\bigg(1 + \frac{\Pi^{\alpha \alpha}_{>>} - \Pi^{\alpha \alpha}_{><}}{\frac{1}{\lambda} + \Pi^{\alpha \alpha}_{><}}\bigg)
\ee
Below we evaluate the differences of polarization functions $\Pi^{\alpha \alpha}_{>>} - \Pi^{\alpha \alpha}_{><}$, and find that different modes, L and T, yield contributions of opposite sign. 

In particular, we find that $\Pi^{33}_{>>} - \Pi^{33}_{><}$ is positive, i.e. the L-modes favor the (2,2) state. This effect of longitudinal modes is well known in the topological insulator literature \cite{Shou-Cheng}. In contrast, the differences $\Pi^{\alpha \alpha}_{>>} - \Pi^{\alpha \alpha}_{><}$ with $\alpha = 1,\,2$ are negative. Thus, the T-modes, which are unique to BLG, favor the (4,0) state. We evaluate Eq.(\ref{eq: deltaf}), and find that the T-modes dominate the free energy, favoring the QAH state.

To proceed with the analysis of 
the quantities $\Pi^{\alpha \alpha}_{>>} - \Pi^{\alpha \alpha}_{><}$, 
it is convenient to define $\epsilon_{\pm} = \epsilon \pm \omega/2$, and $z_{\pm} = |\vec{p} \pm \frac12 \vec{q}|^2/2m$. In this compact notation, we have 
\bea\label{eq: pidiff}
\Pi^{\alpha \alpha}_{>>} - \Pi^{\alpha \alpha}_{><} \!\!&=&\!\! \int 
\frac{F_{\alpha\alpha}(\epsilon,\vec p)}{(\epsilon_+^2 + z_+^2 + \Delta^2)(\epsilon_-^2 + z_-^2 + \Delta^2)} 
,
\\\nonumber
F_{\alpha\alpha}(\epsilon,\vec p)&=&\Delta^2 \Tr (\tau_{\alpha}\tau_3 \tau_{\alpha} \tau_3) + \Delta \Tr (\tau_{\alpha} \tau_{\alpha} \tau_3) (i\epsilon_+) 
\\ \nonumber
&& + \Delta \Tr (\tau_{\alpha}H_0(\vec{p} + \frac12 \vec{q}) \tau_{\alpha}\tau_3),
\eea
where $\int...=\int \frac{d^2p d\epsilon}{(2\pi)^3}...$.
%
Terms in Eq.(\ref{eq: pidiff}) linear in $\Delta$ must vanish, since the fluctuation energy should be invariant under sign changing $\Delta \rightarrow - \Delta$. Technically, the vanishing of terms linear in $\Delta$ follows because $\Tr (\tau_{\alpha} \tau_3 \tau_{\alpha})= 0$, and $\Tr(\tau_{\alpha} H_0 \tau_{\alpha} \tau_3) = 0$. As a result, the first term in $F_{\alpha\alpha}(\epsilon,\vec p)$ (at order $\Delta^2$) is the only term that survives. We can substitute the expression in Eq.(\ref{eq: pidiff}) into Eq.(\ref{eq: deltaf}) and expand the logarithm in small $\Delta^2$, to obtain
%
\begin{eqnarray}\nonumber 
\delta F &=& 4 \int 
\sum_{\alpha} 
\frac{\Delta^2 \rm{Tr}(\tau_{\alpha} \tau_3 \tau_{\alpha} \tau_3)}{D_{\alpha \alpha}(\omega, \vec{q})(\epsilon_+^2 + z_+^2 + \Delta^2)(\epsilon_-^2 + z_-^2 + \Delta^2)}\\
&=& 4 \int 
\left(- \frac{1}{D_{11}(\omega, \vec{q})} - \frac{1}{D_{22}(\omega, \vec{q})} \right.
\\
&& \left. + \frac{1}{D_{33}(\omega, \vec{q})} \right) 
\frac{2 \Delta^2}{(\epsilon_+^2 + z_+^2 + \Delta^2)(\epsilon_-^2 + z_-^2 + \Delta^2)},
\end{eqnarray}
%
where $\int...=\int \frac{d\epsilon d\omega d^2p d^2q}{(2\pi)^6}...$ and 
$D_{\alpha \alpha}(\omega, \vec{q}) = \frac{1}{\lambda} + \Pi^{\alpha \alpha}_{><} (\omega, \vec{q})$.
The integral over $\epsilon$ may be performed exactly by the method of residues, to give
\bea\nonumber
\delta F &=& 4 \Delta^2 \int \frac{d\omega d^2q d^2 p}{(2\pi)^5} \left( - \frac{1}{D_{11}(\omega, \vec{q})} - \frac{1}{D_{22}(\omega, \vec{q})} \right.
\\\nonumber
&& \left. + \frac{1}{D_{33}(\omega, \vec{q})} \right) \frac{\frac{1}{\xi_+} + \frac{1}{\xi_-}}{\omega^2 + (\xi_+ + \xi_-)^2},
\eea
where $\xi_{\pm} = \sqrt{z_{\pm}^2 + \Delta^2}$.
The integral over $p$ may now be performed with logarithmic accuracy. The dominant contributions come from $\xi_{\pm} \approx  0$, and may be evaluated as 
\bea\nonumber
\delta F &=& 8 \Delta^2 \nu_0 \int \frac{d\omega d^2q}{(2\pi)^3} \left(- \frac{1}{D_{11}(\omega, \vec{q})} - \frac{1}{D_{22}(\omega, \vec{q})} \right.
\\
&&\left. + \frac{1}{D_{33}(\omega, \vec{q})} \right) \frac{\ln (r/\Delta)}{r^2}
, 
\label{eq: 13}
\eea
where we have used the pseudo-polar coordinates $r^2 = \omega^2 + (q^2/2m)^2$ and have assumed that $r \gg \Delta$. 

We now have to calculate the various functions $D_{\alpha \alpha}$. We will calculate these quantities analytically with logarithmic accuracy. We begin with the definition $D_{\alpha \alpha} = \frac{1}{\lambda} + \Pi^{\alpha \alpha}_{><}$, where the polarization functions are defined in Eqs.(\ref{eq: pi>>}),(\ref{eq: pi><}). 
We note that the polarization functions $\Pi^{\alpha \alpha}_{ij}$ are logarithmically divergent at small $\omega$, small $|\vec{q}|^2/2m$ and $\Delta=0$. The coefficient of the logarithm can be extracted by setting $\omega, \vec{q}, \Delta = 0$ in the integral in Eqs.(\ref{eq: pi>>}),(\ref{eq: pi><}), 
and introducing an IR cutoff $r$, where $r^2 = \omega^2 + \big(|\vec{q}|^2/2m\big)^2$, and we assume $r\gtrsim\Delta$. In this manner, we obtain
\bea
&& D_{11} = D_{22} = \frac{1}{\lambda} - \int_r^{\Lambda} \frac{d\epsilon d^2p}{(2\pi)^3} \frac{\epsilon^2 - z^2 e^{2i\theta_{\vec{p}}}}{(\epsilon^2 + z^2)^2},
,\quad  
\\
&& D_{33} = \frac{1}{\lambda} - \int_r^{\Lambda} \frac{d\epsilon d^2p}{(2\pi)^3} \frac{\epsilon^2 + z^2}{(\epsilon^2 + z^2)^2}
,
\eea
where we have introduced the notation $ z = \frac{|\vec{p}|^2}{2m}$ and $|\vec{p}|e^{i\theta_{\vec{p}}} = p_x + i p_y$. The integrals may be straightforwardly performed by changing to the pseudopolar coordinates $(\rho, \phi, \theta_p)$, where $\epsilon = \rho \cos \phi$, $z = \rho \sin \phi$, and $\theta_p$ was defined above. The integral goes over $0<\theta_p<2\pi$, $0<\phi<\pi$ and $r<\rho<\Lambda$. Integrating in turn over $\theta_p$, $\phi$ and $\rho$, we find
\begin{eqnarray}
D_{11} = D_{22} = \frac{1}{\lambda} - \frac{\nu_0}{4} \ln \frac{\Lambda}{r}, \quad D_{33} = \frac{1}{\lambda} - \frac{\nu_0}{2} \ln \frac{\Lambda}{r}
.
\end{eqnarray}
We now recall the relation $\lambda^{-1} = \frac12 \nu_0 \ln \Lambda/\Delta$ (the gap equation), and substitute it into Eq.(\ref{eq: 13}), to obtain
\begin{equation}
\delta F = 8 \Delta^2 \nu_0 \int_{\Delta}^{\Lambda} \frac{dr}{r} \left(- \frac{4}{\ln \frac{\Lambda}{\Delta} + \ln \frac{r}{\Delta}} + \frac{1}{\ln \frac{r}{\Delta}} \right) \ln \frac{r}{\Delta}
.
\end{equation}
This integral can be evaluated using the substitution $x = \ln \frac{r}{\Delta}$, giving
\be
\delta F = 8 \Delta^2 \nu_0 \int_{0}^{\ln(\Lambda/\Delta)} \frac{\ln(\Lambda/\Delta)-3x}{\ln(\Lambda/\Delta)+x} dx
.
\ee
Evaluating the integral, we obtain a negative value
\be
\delta F =  8(-3 + 4 \ln2) \Delta^2 \nu_0 \ln \frac{\Lambda}{\Delta} \approx -1.82 \Delta^2 \nu_0 \ln \frac{\Lambda}{\Delta}
, 
\label{eq: result}
\ee
which favors the QAH state. 


It should be noted that the difference in energies between the (4,0) and (2,2) manifolds is of the same order as the mean field energy, so the mean field plus fluctuations analysis is ill controlled. However, it provides us with an intuition about the splitting between manifolds of different signatures, and we believe the qualitative details of the fluctuation splitting are reproduced correctly by this analysis.

We note that our fluctuation analysis included only those modes that correspond to weak coupling instability in BLG. 
We could also have included Stoner modes in our fluctuation analysis. These would produce an additional contribution 
\bea
&& \delta F_{\rm Stoner} = 8 \Delta^2 \nu_0 \int \frac{d\omega d^2q}{(2\pi)^3}\frac{1}{D_{00}(\omega, \vec{q})} \frac{\ln(r/\Delta)}{r^2}
,
\\
&& D_{00} = \frac{1}{\lambda} + \Pi_{><}^{00}
,
\eea
where $\Pi^{00}_{><}$ is defined by Eq.(\ref{eq: pi><}) 
with $\alpha=\beta=0$, i.e. with $\tau_\alpha=\tau_\beta = 1$. Now, since $\Pi^{00}_{><}$ is not log divergent, we can take $D_{00} = 1/\lambda$ with logarithmic accuracy. We then obtain a contribution $\delta F_{\rm Stoner} = 4 \Delta^2 \nu_0 \ln \Lambda/\Delta$, which is sufficiently large to change the sign of the result Eq.(\ref{eq: result}). However, this calculation, which neglects correlation effects, is likely to strongly overestimate the effect of Stoner modes, and therefore we believe that Stoner modes should be left out of the fluctuation analysis.


%

\section{Lifting the degeneracy: thermal fluctuations}

Thermal fluctuations are dominated by gapless Goldstone modes, which are present only in the states that break SU(4) symmetry. In a state ($M_>$, $M_<$), there are $M_> M_<$ Goldstone modes. Thermal fluctuations due to Goldstone modes allow a state to gain entropy, and since the (2,2) states have the most Goldstone modes, they have the highest entropy. It may thus be expected that the (2,2) states dominate at sufficiently high temperature. 



Below we present an analysis showing that this expectation is correct. Since gapless fluctuation modes appear only in the L-mode channel $\delta h\propto \tau_3\delta Q$, it is sufficient to restrict our attention to the L-modes. The general expression for the fluctuation part of the free energy, taking into account L-modes only, is given by a sum over Matsubara frequencies,
\be\label{eq:F_2_goldstone}
F_{\rm fluct}= \frac12 T\sum_{\omega_n,\vec k}\sum_{i,j}\ln \lp \frac1{\lambda}+\Pi^{33}_{ij}(\omega_n,\vec k)\rp
,
\ee
where $\omega_n=2 \pi n T$. 

We will perform a long wavelength expansion of $\Pi^{33}_{ij}(\omega,\vec k)$.
At zeroth order, we note that at $\omega,\,\vec k=0$ the values of $\Pi^{33}_{><}$ and $\Pi^{33}_{>>}$ are given by
%
\bea 
&&\Pi^{33}_{>>}(\omega,\vec k=0) = \frac12 \int\frac{d^2pd\epsilon}{(2\pi)^3}
\tr \lp \tau_3 G_> \tau_3 G_>\rp
\nonumber \\
&&=-
\int \frac{\epsilon^2+z^2 - \Delta^2}{(\epsilon^2+z^2+\Delta^2)^2}
 \frac{d^2pd\epsilon}{(2\pi)^3},
\nonumber\\\nonumber
&& \Pi^{33}_{><}(\omega,\vec k=0) = \frac12 \int \frac{d^2pd\epsilon}{(2\pi)^3}
\tr \lp \tau_3 G_> \tau_3 G_<\rp
\\
&& =-
\int \frac{\epsilon^2+z^2+ \Delta^2}{(\epsilon^2+z^2+\Delta^2)^2}
 \frac{d^2pd\epsilon}{(2\pi)^3}
,
\nonumber
\eea
%
where $G_{>(<)}=1/(i\epsilon-H_0(\vec p)\mp \Delta\tau_3)$.
To distinguish Goldstone modes from gapped modes, it is convenient to recall the gap equation
\begin{equation}
\frac{1}{\lambda} = \int \frac{1}{\epsilon^2+z^2+\Delta^2}
 \frac{d^2pd\epsilon}{(2\pi)^3}
.
\end{equation}
Hence, we have $\frac1{\lambda}+\Pi^{33}_{><}(0)=0$, which corresponds to a Goldstone mode, whereas in the case of $\Pi^{33}_{>>}$ we have
\be\label{eq:Pi_++_goldstone}
\frac1{\lambda}+\Pi^{33}_{>>}(0)=\int \frac{2\Delta^2}{(\epsilon^2+z^2+\Delta^2)^2} \frac{d^2pd\epsilon}{(2\pi)^3}
,
\ee
which is manifestly positive. Thus, Goldstone modes exist only in states $(M_>, M_<)$, where $M_> \neq 0$ and $M_< \neq 0$. 

The free energy, Eq.(\ref{eq:F_2_goldstone}), evaluated at leading order in a long wavelength expansion around  $\omega,\vec k=0$, is given by a sum
\bea\label{eq:F_2_twoterms}
&& F_{\rm fluct}=T\sum_{\omega_n,\vec k} M_>M_<\ln(a\omega_n^2+b\vec k^2)
\\
&&+\frac12( M^2_>+M^2_<)\ln(a'\omega_n^2+b'\vec k^2+c)
,
\eea
where the first term is the contribution of the gapless modes (originating from $\Pi^{33}_{><}$), while the second term is the contribution of the gapped modes (originating from  $\Pi^{33}_{>>}$). The coefficients $a$, $a'$, $b$, $b'$ are obtained by Taylor expanding $\Pi^{33}_{ij}(\omega,\vec k)$ in small $\omega$ and $\vec k$, while $c$ is given by Eq.(\ref{eq:Pi_++_goldstone}). 

To simplify the sum over Matsubara frequencies, it is convenient to define the quantity
$f(u)=T\sum_{\omega_n,\vec k} \ln(\omega_n^2+u^2)$.
%
We can evaluate $f(u)$ by first taking the derivative
\[
\frac{df}{du}=T\sum_{\omega_n,\vec k} \frac1{i\omega_n+u}+ {\rm c.c.}=\coth\frac{u}{2T}
,
\]
and then integrating it over $u$ to obtain 
\[
f(u)=2T\ln\sinh\frac{u}{2T}=2T\ln\lp 1-e^{-u/T}\rp +u-(2\ln 2) T
.
\]
Plugging this identity into the sum (\ref{eq:F_2_twoterms}), we see that the contribution of the gapped modes is exponentially small at low temperatures, $T\ll \sqrt{c/a'}\sim\Delta$, while the sum over gapless modes gives a negative contribution of a power law form,
\be
F_{\rm fluct}=\sum_{\vec k} 2 M_>M_<T\ln \lp 1-e^{-v|\vec q|/T}\rp 
, 
\ee
where $v=b/a\sim \sqrt{\Delta/m}$.
Evaluating the integral, we obtain an estimate
\be\label{eq:F_2_estimate}
F_{\rm fluct}\sim - M_>M_< (\nu_0/\Delta) T^3
,
\ee
which describes the free energy gain due to thermal fluctuations of Goldstone modes.

We see that the gapless Goldstone modes dominate the finite-temperature fluctuation contribution to the free energy. These modes lower the free energy (by increasing entropy). Since the number of gapless modes $M_>M_<$ is maximal for the (2,2) states, these states are entropically favored by thermal fluctuations. 

What is the outcome of competition between the zero-point fluctuations and thermal fluctuations? In Sec.\ref{sec: zero-point} we found that at zero temperature the (0,4) QAH state is energetically favored by zero point fluctuations of the modes ``softened'' under RG. At the same time, the zero-point fluctuations of other modes, such as the Stoner modes, may have an opposite effect, favoring the (2,2) state. In the event the zero-point fluctuation energy is dominated by such non-soft modes, the (2,2) state will be realized in the entire temperature interval where the system is unstable to gap formation.

A more interesting situation may arise if the zero-point fluctuation energy is dominated by the RG-softened modes, favoring the QAH state at zero temperature. In this case, given the opposite effect of zero-point and thermal fluctuations, we have to consider the competition between the QAH and (2,2) states. Since the thermal fluctuation energy (\ref{eq:F_2_estimate}) vanishes at $T=0$, we expect that zero point fluctuations will dominate below a certain temperature $T_*$, above which thermal fluctuations will dominate. If $T_* < T_c$, where $T_c\approx \Delta(T=0)$ is the critical temperature for gap opening, then a QAH state will be realized at low temperatures $0<T<T_*$, whereas a (2,2) gapped state will be realized in the interval $T_*< T< T_c$. In contrast, if $T_* > T_c$, then the QAH state will transition directly to an ungapped state at $T=T_c$ via a second order phase transition, and the (2,2) state will not be realized. 

A rough estimate of the temperature $T_*$ can be obtained by comparing the free energies (\ref{eq:F_2_estimate}) and (\ref{eq: result}),
\be
\delta F_{\rm fluct, (2,2)}\sim - 4\nu_0 \frac{T^3}{\Delta}
,\quad 
\delta F_{\rm fluct, (0,4)}\sim -1.82 \nu_0\Delta^2\ln\frac{\Lambda}{\Delta}
,
\ee
indicating that the scale for $T_*$ is comparable to the temperature $T_c$ at which the gapped state forms. A more detailed analysis of temperature-driven transition between the QAH state and (2,2) state is beyond the scope of this work.



\section{Experimental signatures of the QAH state}

We now discuss experimental tests of the QAH state. The clearest experimental signature
would be detection of the quantum Hall effect at zero external magnetic field. However, detection of this effect requires four probe measurements performed on a sample of BLG that is sufficiently clean and at sufficiently low temperatures as to exhibit spontaneous gap opening \cite{Nandkishore}. Such measurements have not yet been performed. Moreover, detection of this effect could be complicated by the formation of domains with opposite signs of $\Delta$. 
Different domains will have opposite $\sigma_{xy}$, so the Hall conductance of a macroscopic sample will average to a value near zero. However, if there is percolation of edges, there will be a non-vanishing two-terminal conductance of order $e^2/h$.

 Alternative experimental tests of the QAH state may be performed by examining the electronic compressibility in weak magnetic fields. When the chemical potential sits near the missing Landau level in Fig.(\ref{fig: phase diagram}), there should be a gap that extrapolates to a non-zero value as $B \rightarrow 0$. This effect will be seen at either $\nu = 4$ or $\nu = -4$ if there is only one domain, and at $\nu = \pm 4$ if there are multiple domains. 
 
 The gap at $\nu = 4$ will be strengthened by the mechanism outlined around Eq.(\ref{eq: mechanism}), however, a signal at $\nu = -4$ will be seen only if the QAH state is intrinsic, rather than field induced. An incompressible region at $\nu = -4$ combined with a gapped state at $B=0$ can thus be taken as a diagnostic for a QAH state at $B=0$. The filling factors $\nu = \pm 4$ are not equivalent because the QAH state breaks particle-hole symmetry in magnetic field.
 
 Another experimental signature is a phase transition at filling factor $\nu = 0$ and finite $B$ from a QAH state to the Quantum Hall Ferromagnet (QHF) states that are expected to form at large magnetic fields \cite{Barlas}. Such a phase transition would not be seen if the dominant state at small $B$ was of (2,2) type, since the (2,2) states are smoothly connected to the QHF state. 
 
 An incompressible region at $\nu = \pm4$ that occurs at anomalously low magnetic fields, such that the features in compressibility at other integer $\nu$ values are washed out, was found in recent experiments that employed a capacitance scanning probe to study suspended BLG samples \cite{Feldman2}. In transport measurements \cite{Weitz} performed on the same system, a state with finite two-terminal conductance of order $e^2/h$ was found at zero field, which at a finite $B$ field undergoes a transition to an insulating state. These measurements are all compatible with the QAH state, however, since there is as yet no four-terminal measurement, it is not possible to say for certain whether a QAH state has been observed. 

In summary, our symmetry classification of 
the various gapped states proposed for BLG singles out the QAH state as the only gapped state not breaking any continuous symmetry. 
We have investigated the fluctuation-induced splitting of the gapped states, and concluded
that at zero temperature and zero field, the leading instability is to the QAH state. We have discussed the phenomenology and experimental signatures of this state, and have shown that it can be stabilized by weak external magnetic field.

We thank M. Allen, B. Feldman, J. Martin, T. Weitz, and A. Yacoby for sharing with us unpublished data.
We also acknowledge useful discussions with V. Deshpande, A. Young, S. E. Korshunov and P.Kim. This work was supported by Office of Naval Research Grant No. N00014-09-1-0724.

\end{document}